\documentclass[fleqn,usenatbib]{mnras}

\usepackage{newtxtext,newtxmath}

\usepackage[T1]{fontenc}
\usepackage{ae,aecompl}

\usepackage{graphicx}	
\usepackage{amsmath}	
\usepackage{amssymb}	
\usepackage[usenames]{xcolor}
\usepackage{color}
\usepackage{soul}

\title[An unusually large gaseous transit]{An unusually large gaseous transit in a debris disc}

\author[D. P. Iglesias et al.]{
Daniela P. Iglesias$^{1,2}$\thanks{E-mail: dpiglesi@puc.cl},
Johan Olofsson$^{1,2}$, 
Amelia Bayo$^{1,2}$, 
Sebastian Zieba$^{3}$,
\newauthor
Mat\'ias Montesinos$^{1,2,4}$,
Jonathan Smoker$^{5}$,
Grant M. Kennedy$^{6}$,
Nicol\'as Godoy$^{1,2}$,
\newauthor
Blake Pantoja$^{7}$,
Geert Jan Talens$^{8}$,
Zahed Wahhaj$^{5}$,
and Catalina Zamora$^{1,2}$
            \\
            \\
$^{1}$Instituto  de  F\'isica  y  Astronom\'ia,  Facultad  de  Ciencias,  Universidad de Valpara\'iso, Av. Gran Breta\~na 1111, 5030 Casilla, Valpara\'iso, Chile\\
$^{2}$N\'ucleo Milenio de Formaci\'on Planetaria - NPF, Universidad de Valpara\'iso, Av. Gran Breta\~na 1111, Valpara\'iso, Chile\\
$^{3}$Universit\"at Innsbruck, Institut f\"ur Astro- und Teilchenphysik, Technikerstra{\ss}e 25, 6020 Innsbruck, Austria \\
$^{4}$Chinese Academy of Sciences South America Center for Astronomy, National Astronomical Observatories, CAS, Beijing 100012, China,\\
$^{5}$European Southern Observatory, Alonso de C\'ordova 3107, Vitacura, Santiago, Chile \\
$^{6}$Department of Physics, University of Warwick, Gibbet Hill Road, Coventry CV4 7AL, UK \\
$^{7}$Departamento de Astronom\'ia, Universidad de Chile, Camino el Observatorio 1515, Las Condes, Santiago, Chile, Casilla 36-D\\
$^{8}$Institut de Recherche sur les Exoplan\`etes, D\'epartement de Physique, Universit\'e de Montr\'eal, Montr\'eal, QC H3C 3J7, Canada\\
}

\date{Accepted XXX. Received YYY; in original form ZZZ}

\pubyear{2019}

\begin{document}
\label{firstpage}
\pagerange{\pageref{firstpage}--\pageref{lastpage}}
\maketitle

\begin{abstract}

We present the detection of an unusually large transient gas absorption in several ionized species in the debris disc star HD\,37306 using high-resolution optical spectra. We have been analysing a large sample of debris discs searching for circumstellar gas absorptions aiming to determine the frequency of gas in debris discs. 
HD\,37306 stood out showing remarkably broad absorptions superimposed onto several photospheric Ca\,{\sc ii}, Fe\,{\sc ii} and Ti\,{\sc ii} lines. The observed absorptions, unlike typical exocometary transits, lasted for at least eight days. Here we analyse simultaneous spectroscopic and photometric data of the event and evaluate different scenarios that might explain the observed features. We conclude that the most likely scenario might be an exocometary break-up releasing a significant amount of gas close to the star, producing an occulting ``ring"/``torus" shape.
\end{abstract}

\begin{keywords}
circumstellar matter -- stars: individual: HD\,37306 -- comets: general -- planetary systems.
\end{keywords}



\section{Introduction}


Debris discs were long thought to be second generation dusty discs completely devoid of gas. However, this paradigm has changed in the last few years with the detection of gas in a growing number of young debris discs (e.g. \citealt{Kospal2013}, \citealt{Moor2015}, \citealt{Rebollido2018} and the references in \citealt{Kral2017}).
In our ongoing survey to robustly estimate the fraction of debris discs harboring circumstellar gas (described in \citealt{Iglesias2018}), we have followed the methodology described in \citealt{Kiefer2014HD172555}. In short, we searched for narrow absorption features superimposed onto the photospheric absorption lines in the Ca\,{\sc ii} doublet H \& K. These absorption features, when variable, can be interpreted as arising from comets falling towards the star (hence the term Falling Evaporating Bodies, FEBs).

Our survey sample consists of $\sim$300 objects selected with a methodology that is unbiased in terms of disc inclination. We have gathered several epochs of observations for $\sim$91\% of them, compiling a rich database of thousands of spectra. The full variability analysis will be presented in Iglesias et al. (in prep), 
but in this work, we focus on a particularly intriguing object: HD\,37306.

HD\,37306, a bright A1V-type star, was previously studied by our team in \cite{Iglesias2018}. With the data available at the time, we concluded that the two stable ``extra" components in the Ca\,{\sc ii} lines that the object presented, were of interstellar origin. In that study we analysed 26 high-resolution observations from February 2006 to March 2016 where no variability was detected. Recently, we updated our database collecting new publicly available observations of our sample from the ESO archive. We found eight new observations of HD\,37306 taken in 2017 where remarkable additional absorption features in several metallic lines (also reported in Rebollido et al. submitted) were detected, particularly large in the Ca\,{\sc ii} lines. In this study we complement this data with additional spectroscopy and time series photometry and present different scenarios to explain the origin of these transient features.

\section{HD 37306: Stellar parameters and Spectral Energy Distribution (SED)} %

In \cite{Iglesias2018} we estimated some of the stellar properties by fitting Kurucz models \citep{Castelli1997} to the  Ca\,{\sc ii} (at 3933.66 \& 3968.47 \AA) and Na\,{\sc i} (at 5889.95 \& 5895.92 \AA) doublets. We estimated a $v\sin{i}$ of $140\pm5$\,km.s$^{-1}$ (consistent with previous estimates), a $T_{\mathrm{eff}}$ of $8800\pm50$ K (in agreement with Gaia DR2 estimate of $9138\pm_{386}^{297}$ K, \citealt{GaiaDR2_2018}), $\log g$ of $4.15\pm0.38$ {\it dex} and {\it [Fe/H]} of $-0.12\pm0.22$. We also estimated a heliocentric radial velocity of $23.49\pm1.28$\,km.s$^{-1}$, consistent with the previous measurement found in the literature of $23.00\pm0.70$\,km.s$^{-1}$ \citep{Gontcharov2006}. HD\,37306 is located at a distance of $70.46\pm0.39$\,pc \citep{GaiaDR2_2018} and it has been reported to be a member of the Columba Association \citep{Zuckerman_Song2012}. We assessed the probability of belonging to the Columba Association with the {\sc banyan} $\Sigma$\footnote{\url{http://www.exoplanetes.umontreal.ca/banyan/banyansigma.php}} (Bayesian Analysis for Nearby Young AssociatioNs $\Sigma$) tool \citep{Gagne2018} and confirmed its membership with a 99.3\% probability. Incidentally, an isochronal age of $28.53\pm^{307.72}_{28.52}$ Myrs was estimated for the object using VOSA\footnote{\url{http://svo2.cab.inta-csic.es/theory/vosa/}} \citep{Bayo2008} based on the SED fitted parameters and different sets of isochrones (\citealt{Baraffe1998}, \citealt{Siess2000}). Thus we adopt the age of 30 Myrs of the Columba Association \citep{Torres2006}. Ages from the literature range between 10 Myrs (e.g. \citealt{Ballering13}, \citealt{DeRosa14}, \citealt{David15}) and 453 Myrs \citep{Gontcharov12}. 

\begin{figure}
	\includegraphics[width=\columnwidth]{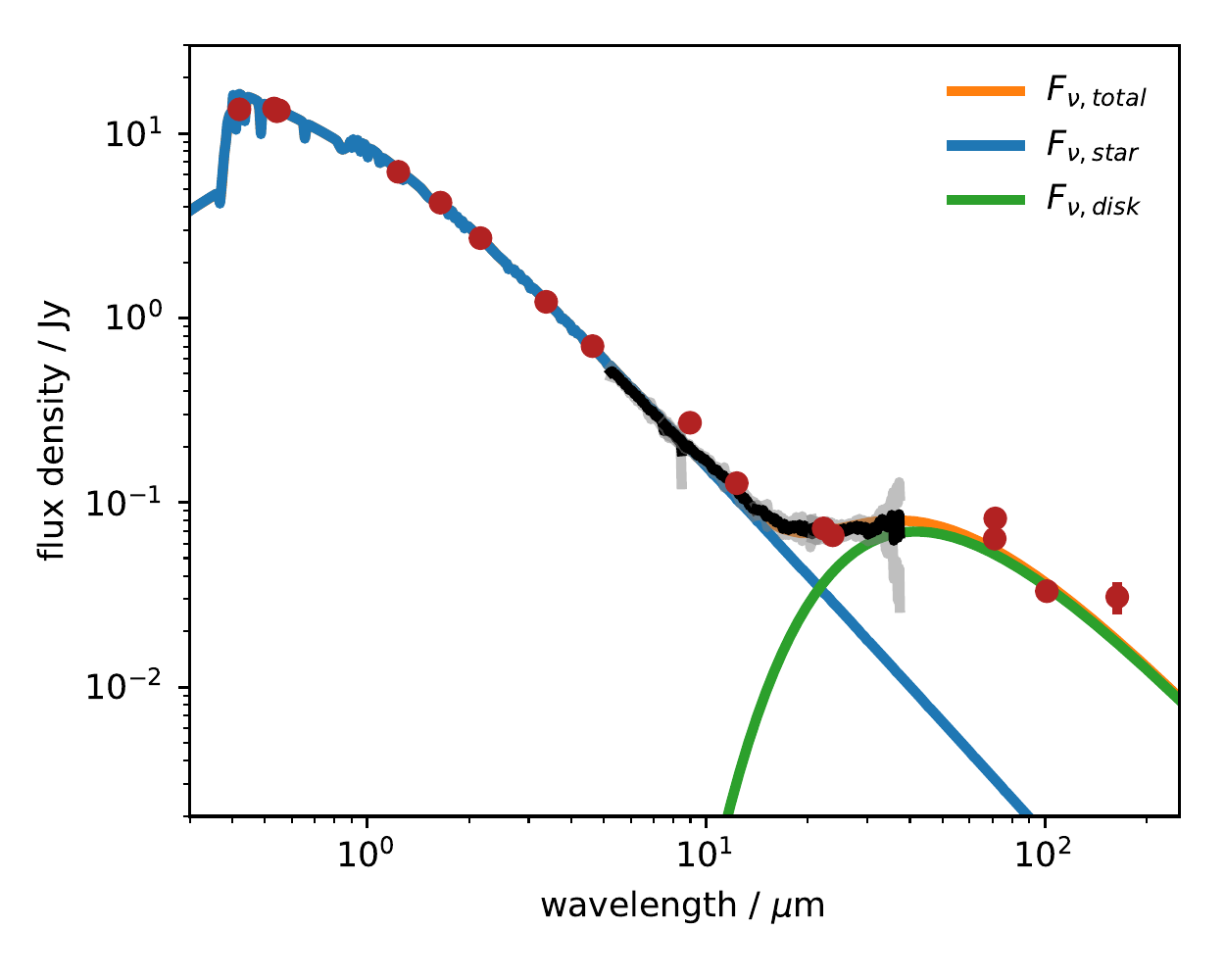}
    \caption{Flux distribution for HD\,37306. Dots show photometry and the black and grey lines show the \emph{Spitzer} IRS spectroscopy  and  uncertainty. The blue line shows the best-fit stellar photosphere, and the green line the best-fit modified blackbody for the disc. }
    \label{fig:SED}
\end{figure}

Regarding its known local properties, HD\,37306 is surrounded by a debris disc \citep{Zuckerman_Song2012}, confirmed by the excess emission in the SED (see Fig. \ref{fig:SED}). We have fit model grids for the star and disc to the data. We used synthetic photometry of the models to fit the photometry, and resampled model spectra to fit the IRS spectrum. The model is composed of a PHOENIX model atmosphere \citep{Husser2013} at a given temperature, which is normalized to the optical photometry by solid angle, and a model for the thermal emission depending on the temperature of the dust, where the Planck function is normalized to the mid-/far-infrared emission by, basically, the dust mass in the disc. 
The  best  fitting  model parameters are found with the \emph{MultiNest} code \citep{Feroz2009}, with both the stellar and disc parameters found simultaneously. 
The best fit model has a stellar $T_{\rm eff} = 9100$K, consistent with the previous estimates. The dust component has a temperature of $120$K and fractional luminosity $L_{\rm disc}/L_\star = (7 \pm 0.2) \times 10^{-5}$. Based on this temperature and the stellar luminosity of $17L_\odot$, the blackbody radius for the dust is 21au. A weak silicate feature is visible in the IRS spectrum when the best fitting model is subtracted, indicating that at least some small ($\sim \mu$m sized) dust is present, and therefore that the dust is not entirely comprised of grains large enough to behave as blackbodies. Thus, the true distance of the dust from the star is likely larger than 21au, probably of the order of 50au.

\section{Observations and data analysis}


\subsection{Spectroscopic Data}
\label{sec:obs} 


We have collected a total of 35 high-resolution (R = 48,000 -- 115,000) optical spectra from different instruments with wavelength coverages within a range $\lambda  \sim$ 3350 -- 9500\AA, combining our own observations and others from the ESO archive. Our data set includes spectra taken with HARPS \citep{Mayor2003} mounted on the ESO 3.6m telescope and FEROS \citep{Kaufer1999} on the MPG/ESO 2.2m telescope, both at La Silla Observatory in Chile, and MIKE \citep{Bernstein2003} mounted on the Magellan-Clay telescope at Las Campanas Observatory, Chile. The dates of the observations, the instruments used and the number of spectra per night are summarized in Table \ref{tab:obs_table}.

All the observations were reduced using the standard pipelines of each instrument. The ESO spectra were downloaded from the Phase 3 portal, and the MIKE spectra were reduced with the CarPy pipeline (\citealt{Kelson2000}, \citealt{Kelson2003}). Barycentric radial velocity corrections were applied to the MIKE spectra as these corrections are not included in its pipeline. Telluric line contamination was removed using {\sc molecfit}\footnote{\url{http://www.eso.org/sci/software/pipelines/skytools/molecfit}} (\citealt{Smette2015}, \citealt{Kausch2015}) in the same way as in \cite{Iglesias2018}. 

\begin{figure}
	\includegraphics[width=\columnwidth]{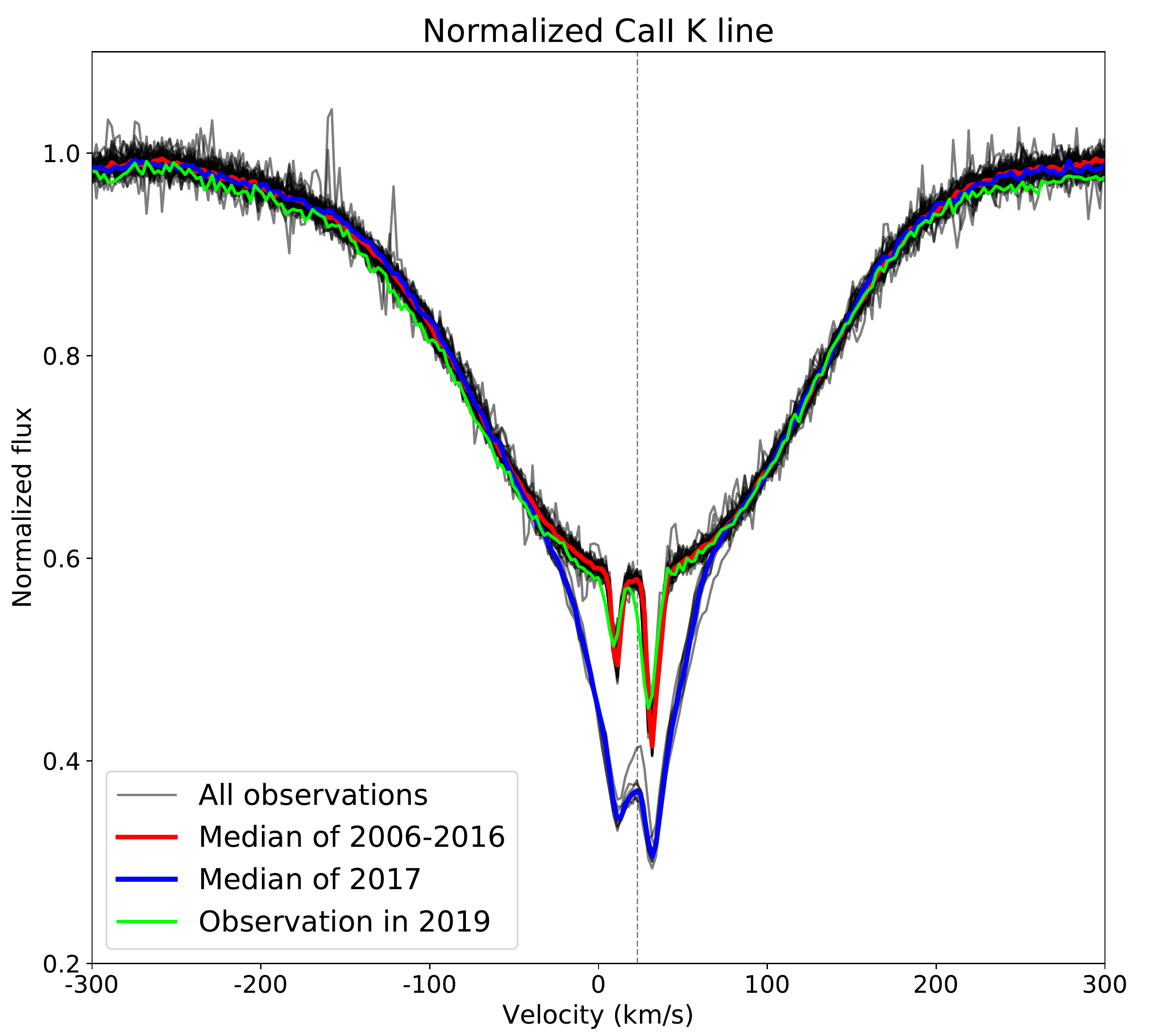}
    \caption{All the Ca\,{\sc ii} K line observations of HD\,37306 normalized and over-plotted for comparison. The median spectra of all the epochs before and during the event are shown in red and blue, respectively. The spectrum taken in 2019-03-18 is highlighted in green. The stellar radial velocity is marked with a dashed line. }
    \label{fig:CaIIK}
\end{figure}

In order to quantify the extra absorption features detected in September 2017, we normalized each spectral line and computed a median from all the observations without the absorption along with another median from all the observations where the additional absorption was detected. An example for the Ca\,{\sc ii} K line can be seen in Fig. \ref{fig:CaIIK}. Then, we divided the median containing the absorption by the median with no detection to analyse the properties of the feature isolated from the photospheric line and additional interstellar features.

\begin{figure}
	\includegraphics[width=\columnwidth]{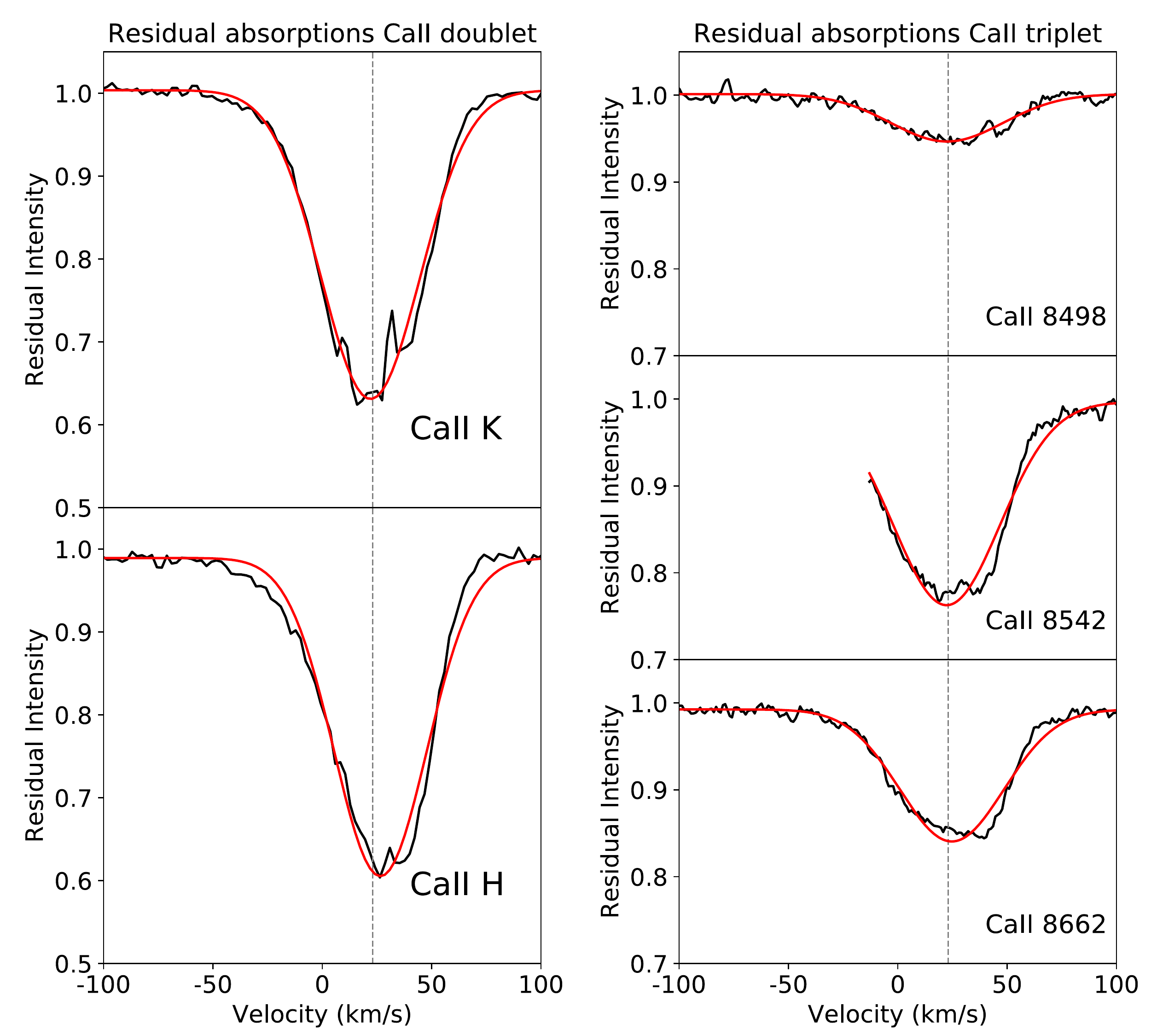}
    \caption{Median of the residuals of the Ca\,{\sc ii} doublet and triplet of HD\,37306. Gaussian fits in red. The Ca\,{\sc ii} line at 8542\AA\ is truncated due to an instrumental gap. Again, the radial velocity of the star is marked with a gray dashed line. }
    \label{fig:allCaII}
\end{figure}

\begin{figure}
	\includegraphics[width=\columnwidth]{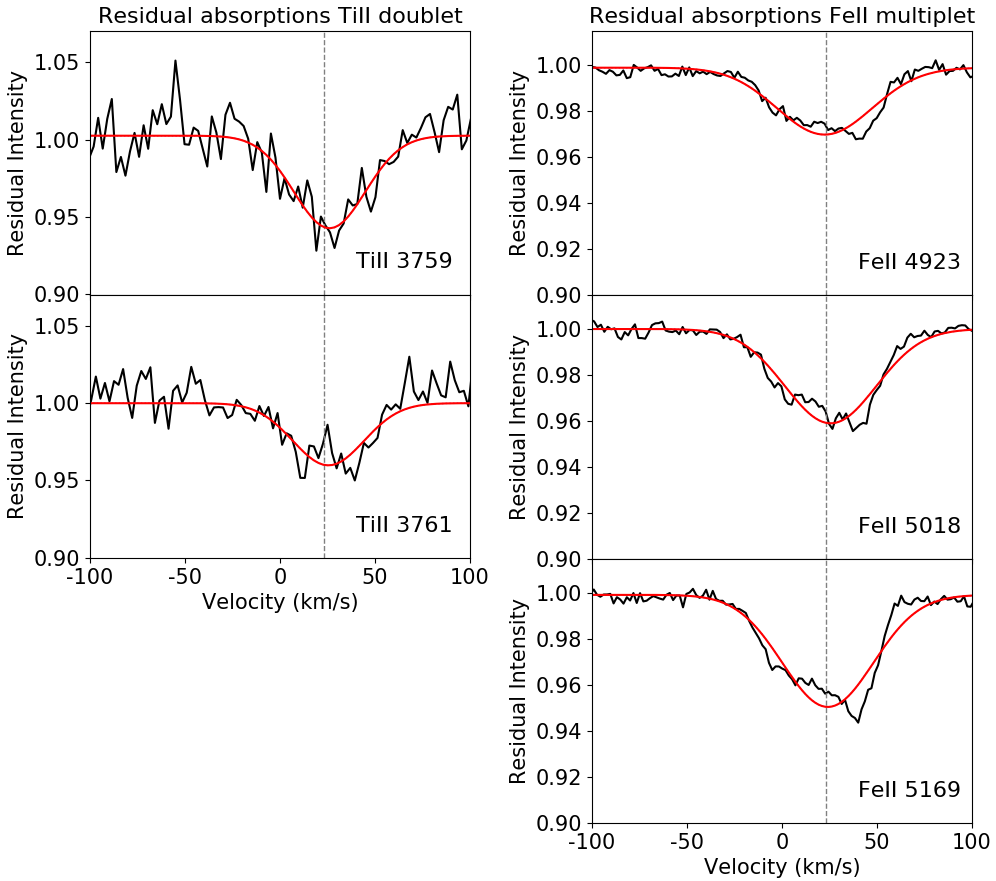}
    \caption{Median of the residuals of the Ti\,{\sc ii} doublet and Fe\,{\sc ii} multiplet of HD\,37306. }
    \label{fig:TiIIFeII}
\end{figure}

We consider as a ``detection'' those absorptions in the residual spectrum (resulting from the division of the medians) that exceed $3\sigma$. As these residual features are unusually wide, and thus, seem to be dominated by Doppler broadening we fit Gaussian profiles to these absorptions in order to estimate their parameters (see Table \ref{tab:params}). Examples of these residual absorptions and their Gaussian fits are shown in Fig. \ref{fig:allCaII} for the Ca\,{\sc ii} lines and Fig. \ref{fig:TiIIFeII} for Ti\,{\sc ii} and Fe\,{\sc ii}.

The observed wavelength of each transition in air values were taken from the NIST Atomic Spectra Database\footnote{\url{https://www.nist.gov/pml/atomic-spectra-database}}. The heliocentric radial velicities of the absorptions were estimated from the center of the Gaussian fits. The apparent column densities $N$ were computed following \citet{Savage1991} using oscillator strength values from \citet{Morton1991} in the case of Ca\,{\sc ii} and from the NIST Atomic Spectra Database for Ti\,{\sc ii} and Fe\,{\sc ii}. Equivalent width and Doppler broadening parameter $b$ were also derived from the Gaussian fit, with $b=FWHM/2\sqrt{\ln{2}}$. The instrumental effects on $b$ are negligible in this case.

\begin{table}
	\centering
	\caption{Spectroscopic data used in this work. Instrument, number of spectra per night and UT dates of each observation.} 
	\label{tab:obs_table}
	\begin{tabular}{lccr} 
		\hline
		Instrument &  N Spectra & Dates \\
		\hline
		HARPS & 4, 2, 2, 2, 2 & 2006-02-[08, 09, 10, 11, 13] \\
		HARPS & 2, 2 & 2006-03-12, 2006-11-18 \\
		HARPS & 2, 2, 2 & 2007-12-[05, 06, 10] \\
		FEROS & 2 & 2015-10-23 \\
		FEROS & 1, 1 & 2016-03-[28, 29] \\
		FEROS & 2, 1, 2, 1, 1, 1 & 2017-09-[23, 24, 25, 26, 27, 30] \\
		MIKE & 1 & 2019-03-18 \\
		\hline
	\end{tabular}
\end{table}

\begin{table}
	\centering
	\caption{Parameters of the observed absorption lines. Element, observed wavelength of the transition in air, heliocentric radial velocity of the absorption, logarithm of the apparent column density $N$, equivalent width and Doppler broadening parameter $b$. Errors in $V_{\odot}$ are in the order of 2--3 $km\,s^{-1}$ and for $\log_{10} N$, EW and $b$ are in the order of 10--15\%.} 
	\label{tab:params}
	\begin{tabular}{lccccc} 
		\hline
		Elem. &  Wavel. & $V_{\odot}$ & $\log_{10} N$  & EW & $b$\\
		        &    (\AA)  & ($km\,s^{-1}$) & ($cm^{-2}$) & (m\AA) & ($km\,s^{-1}$) \\
		\hline
		Ti\,{\sc ii} & 3759.30 & 26.11 & 12.16 & 35.21 & 26.58 \\
		Ti\,{\sc ii} & 3761.33 & 25.49  & 11.88 & 23.62 & 26.47 \\
		Ca\,{\sc ii} & 3933.66 & 22.00 & 12.57 & 276.11 & 31.95 \\
		Ca\,{\sc ii} & 3968.47 & 26.54 & 12.87 & 273.05 & 29.92 \\
		Fe\,{\sc ii} & 4233.16 & 25.12 & 13.75 & 22.71 & 30.98 \\
		Fe\,{\sc ii} & 4549.47 & 26.85 & 13.82 & 27.63 & 34.75 \\
		Fe\,{\sc ii} & 4555.89 & 18.90 & 14.12 & 17.05 & 46.76 \\
		Fe\,{\sc ii} & 4583.83 & 25.08 & 13.90 & 26.51 & 32.72 \\
		Fe\,{\sc ii} & 4629.33 & 28.17 & 14.04 & 11.50 & 35.16 \\
		Fe\,{\sc ii} & 4923.92 & 22.43 & 13.13 & 30.10 & 35.41 \\
		Fe\,{\sc ii} & 5018.44 & 25.59 & 13.40 & 41.39 & 33.95 \\
		Fe\,{\sc ii} & 5169.03 & 24.14 & 12.98 & 50.60 & 33.88 \\
		Fe\,{\sc ii} & 5197.57 & 25.27 & 13.50 & 11.20 & 28.23 \\
		Fe\,{\sc ii} & 5234.62 & 25.96 & 13.89 & 14.17 & 30.51 \\
		Fe\,{\sc ii} & 5275.99 & 24.76 & 13.62 & 13.20 & 28.24 \\
		Fe\,{\sc ii} & 5316.61 & 28.40 & 13.89 & 26.69 & 30.87 \\
		Fe\,{\sc ii} & 6247.56 & 24.39 & 13.68 & 9.86 & 36.40 \\
		Fe\,{\sc ii} & 6456.38 & 24.43 & 13.67 & 13.64 & 31.25 \\
		Ca\,{\sc ii} & 8498.02 & 22.49 & 13.17 & 99.65 & 36.37 \\
		Ca\,{\sc ii} & 8542.09 & 22.32 & 13.00 & 379.12 & 34.51 \\
		Ca\,{\sc ii} & 8662.14 & 24.72 & 12.90 & 264.05 & 33.64 \\

		\hline
	\end{tabular}
\end{table}

\subsection{Photometric Data}

In order to complement our spectroscopic observations, we mined photometric databases for simultaneous observations. In particular, we collected photometric measurements from MASCARA North (\citealt{Talens2017}, \citealt{Talens2018}), ASAS-SN (\citealt{Shappee2014}, \citealt{Kochanek2017})
and TESS \citep{Ricker2015}.

\subsubsection{Multi-site All-Sky CAmeRA}

\begin{figure}
	\includegraphics[width=\columnwidth]{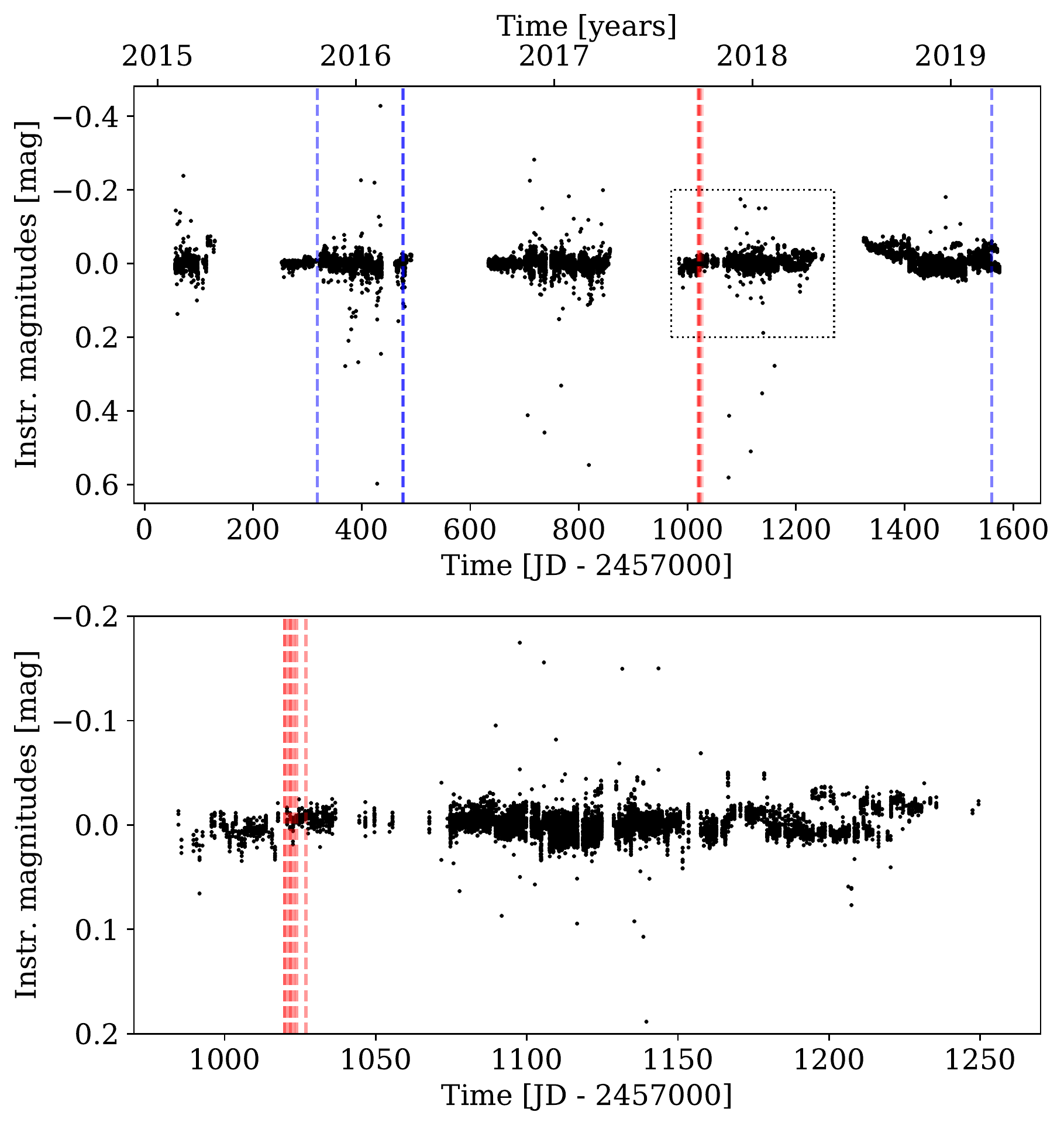}
    \caption{\textit{Upper panel}: The full MASCARA light curve of HD\,37306. The blue and red vertical dashed lines indicate the times of spectroscopic observations as seen in Table \ref{tab:obs_table}. The marks in red correspond to the spectroscopic observations presenting the event. No variations were  detected in spectroscopic observations of the epochs marked in blue. \textit{Lower panel}: Zoom into the black dotted box seen in the upper panel.}
    \label{fig:MASCARA}
\end{figure}

MASCARA North (Multi-site All-Sky CAmeRA; \citealt{Talens2017}, \citealt{Talens2018}), located at the Observatorio del Roque de los Muchachos on La Palma, observed HD\,37306 between February 2015 and April 2019 for several hours each night the star was visible. Photometry was extracted from the 6.4 seconds cadence observations and processed using the primary calibration procedure outlined in \cite{Talens2018}, which uses groups of stars to correct for systematics common to many light curves. This primary calibration corrects for the effects of the variable atmosphere, the camera transmission and intra-pixel variations before binning the data to a cadence of 320 seconds. After binning, a secondary calibration step is used to remove residual systematics in the light curves of individual stars. The fully calibrated light curve is shown in Fig. \ref{fig:MASCARA} (upper panel).

Part of the observations were taken simultaneously with the 2017 spectroscopic campaign, as can be seen in Fig. \ref{fig:MASCARA} (zoom in the lower panel). These observations started mid September 2017. No significant change in flux was observed during or after this time.

\subsubsection{ASAS-SN}

The ``All-Sky Automated Survey for Supernovae'' (ASAS-SN) project (\citealt{Shappee2014}, \citealt{Kochanek2017}) consists of 24 telescopes in different locations around the globe. This survey has taken multiple observations of HD\,37306 from 2013-09-19 up to date. In this case, we took into account the fact that the observations were taken with several different instruments and normalized each light curve per instrument by its median value for better comparison. However, similar to the observations taken with MASCARA, no significant changes were detected in the light curves.

\subsubsection{TESS}
\begin{figure}
	\includegraphics[width=\columnwidth]{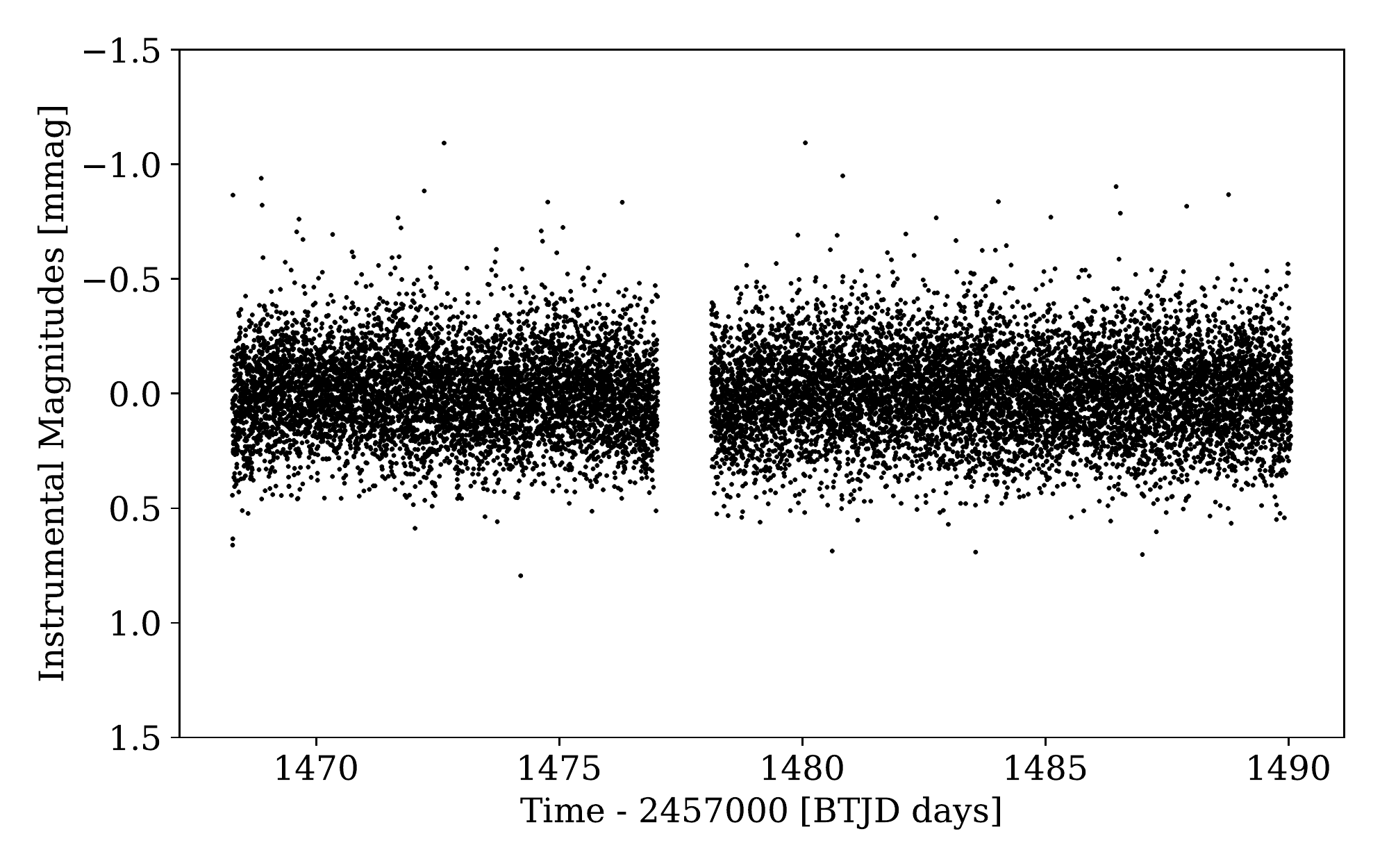}
    \caption{The full TESS light curve of the star HD\,37306. The gap of about one day in the light curve is caused by a pause in observations due to TESS downlinking the data to Earth.}
    \label{fig:TESS}
\end{figure}

The Transiting Exoplanet Survey Satellite (TESS; \citealt{Ricker2015}) observed HD\,37306 in its 6$^{\text{th}}$ Sector from 12 December 2018 to 6 January 2019 (unfortunately not within the spectroscopic transient event). 
HD\,37306 (TIC\,287842651, T = 6.08 mag) is one of the preselected targets for which short cadence (i.e. 2-minute) data is provided. We use the 2-minute Presearch Data Conditioning (PDC; \citealt{Smith2012}; \citealt{Stumpe2012}) light curve from the Science Processing Operations Center (SPOC) pipeline (\citealt{Jenkins2016}; \citealt{Jenkins2017}), which was originally developed for the Kepler mission \citep{Jenkins2010}. These light curves are corrected for systematics by the SPOC pipeline. We also remove every measurement with a non-zero ``quality'' flag (see \S9 in the TESS Science Data Products Description Document\footnote{\url{https://archive.stsci.edu/missions/tess/doc/EXP-TESS-ARC-ICD-TM-0014.pdf}}) which mark anomalies like cosmic ray events or instrumental issues. This finally gives as a time span of 21.8 days and a duty cycle of 95\%. The resulting full light curve can be seen in Figure \ref{fig:TESS}. 

A frequency analysis using the software package {\sc Period04} \citep{Lenz05} that combines Fourier and least-squares algorithms was conducted. It reveals no significant frequencies down to a signal-to-noise ratio of 4 (following the analysis in \citealt{Breger1993}). 

\subsection{Imaging Data}

Additionally, we collected publicly available SPHERE \citep{Beuzit2019} observations of HD\,37306 from the ESO archive (program ID 095.C-0212).
The observations were performed using the dual-band imaging mode IRDIFS ($H2H3$, \citealt{Vigan2010}, \citealt{Dohlen2008}, \citealt{Claudi2008}). Basic data reduction was performed using the SPHERE Data Reduction Handling pipeline \citep{Pavlov2008}, to perform background subtraction, flat-field and bad pixels corrections, and to determine the position of the star behind the coronograph. To try to detect the debris disc we processed the data using the angular differential imaging technique (\citealt{Marois2006}, \citealt{Gomez2017}) by performing a principal component analysis before de-rotating and stacking the pupil-tracking observations. The sky rotation achieved during the observations was of 29.2$^{\circ}$. We removed between $1$ and $25$ principal components, but did not detect the disc in the reduced images. The 5$\sigma$ contrast achieved were in the order of 14.6mag at 1 arcsec and 16.8mag at 3 arcsec. The FOV of 11''x11'' considering the star is located at $\sim$70pc yields a coverage of $\sim(775au)^{2}$. Given the extra absorption detected in the Ca\,{\sc ii}, Fe\,{\sc ii} and Ti\,{\sc ii} lines, if the event is related to a body belonging to the debris disc, the disc is most likely close to edge-on, the most favorable case for a detection using angular differential imaging \citep{Milli2012}. Despite this possible favorable orientation, the non detection of the disc with SPHERE can most likely be explained either with a very small disc close to the star, or a low surface disc brightness (either an intrinsically faint disc or a radially very extended one). 

\section{Transient spectroscopic Event}


In every observation of HD\,37306 taken in September 2017 with the FEROS spectrograph, unusually large additional absorption features were present in the Ca\,{\sc ii} lines and other metallic lines such as Ti\,{\sc ii} and Fe\,{\sc ii}. These observations cover a range of eight consecutive nights  and the absorptions are consistently present during the whole time range. Once we analysed this data-set, on March 2019, we took a new epoch of data with MIKE, and found that the additional absorptions had disappeared, going back to the ``quiescent'' stage. Unfortunately, we cannot determine the duration of the event detected in HD\,37306 because previous and posterior data were taken $\sim$1.5 years apart. Therefore, we can only state that the event lasted at least eight days, but less than three years. A detailed view of the Ca\,{\sc ii} K line is shown in Fig. \ref{fig:CaIIK}. The large additional absorption detected in 2017 can be appreciated in blue, while previous and posterior data only show two narrow features of interstellar origin \citep{Iglesias2018}.

 We analysed other typical gas tracers such as Na\,{\sc i}, O\,{\sc i} and the Balmer lines, not finding any change to our 3$\sigma$ threshold (see Sec.~\ref{sec:obs}). We also studied SiO bands as collision (hyper-velocity impact) released gas would be expected to re-condense, but we observed no change on those bands either. 
 
The absorptions in the H and K lines are equally intense due to saturation, suggesting optically thick material. Based on the absorption depth of the Ca\,{\sc ii} H \& K lines and following the guidelines in \cite{Kiefer2014BetaPicNature}, 
we estimated that the stellar fraction covered by the gas would be $\sim$37\% and that the gas could be roughly located at about 15 stellar radii (or about 0.14 AU for the typical size of an A1V-type star of 2.0 $R_{\odot}$, \citealt{PecautMamajek2013}). The higher temperature closer to the star is consistent with the detection of only ionized species and not neutral, and is also consistent with the feature exhibiting broader components due to thermal broadening. 

The FWHM of the absorption features are in the order of $\sim 50$\,km.s$^{-1}$, much larger than the typical values observed during exocometary transits ($10-15$\,km.s$^{-1}$, \citealt{Kiefer2014BetaPicNature}, \citealt{Beust2001}) and their radial velocities are centered close to the stellar one, as can be seen in Figs. \ref{fig:allCaII} and \ref{fig:TiIIFeII}, and in Table \ref{tab:params}. There are no significant changes either in the intensity of the absorptions or their radial velocities during the September 2017 period. Regarding the duration and stability of this event, although we do not know exactly how long it lasted, a duration $\geq$8 days is remarkably high in the context of exocometary transits, as they usually evolve in less than 1 day (\citealt{Kiefer2014twoAtype}, \citealt{Welsh2018}).  
Considering the characteristics of the features, we expect that the gas producing them would be close to the star, as the radiation pressure on the ions is larger and acts more efficiently at shorter distances. A high radiation pressure could be responsible for a higher velocity dispersion of the ions, leading to broader components \citep{Beust1991}. Recent work by \citet{Lin2019} highlights the possibility that gas released from bodies that are on eccentric orbits, whether they be planetesimals or smaller dust, should also reside on an eccentric orbit. Thus, it is possible that some or all of the gas velocity dispersion is related to its origin, rather than radiation pressure acting solely on the gas. 


\section{Discussion}

We here explore different possibilities that might explain the observed absorptions and evaluate which one could be the most likely scenario. 

\subsection{Instrumental Artifact}

In order to discard the possibility that the absorptions might be due to an instrumental issue, we also analysed spectra from other A-type stars taken during the same nights with the same instrumental set-up. We did not detect any similar event in these other sources, thus firmly ruling out any possible instrumental artifact.

\subsection{Circumstellar Shell}

A possible explanation for the observed spectroscopic event could be a circumstellar shell. This scenario could explain the long duration and stability of the absorptions. In addition, absorption features in the Ti\,{\sc ii} and Fe\,{\sc ii} lines are usually present in shell stars. However, shell stars typically present narrow absorption features at the core of the Hydrogen lines \citep{Slettebak1986,GrayCorbally2009}. As previously mentioned, we analysed the Balmer H lines of HD\,37306, not finding any sign of non-photospheric absorption or emission or any variation at all (to a 3$\sigma$ level) through all the observations. The relatively low projected rotational velocity of HD\,37306 (considering we are discussing fast rotators given the spectral type) of $140$\,km.s$^{-1}$ also argues against this scenario, since, according to \cite{Abt2008} and 
\cite{Abt1997}, evidence of ``shell-like" circumstellar gas is mainly detected for stars with $v\sin i > 200$\,km.s$^{-1}$. Although the origin of circumstellar shells is unclear, some studies suggest they are accreted by rapidly rotating stars ($v\sin i \geq 200$\,km.s$^{-1}$) from the interstellar medium (
\citealt{Abt2015}). In this particular case, it would be unlikely that HD\,37306 accreted and lost such a shell within three years. On the other hand, considering the possibility of the shell being expelled from the star, as it seems to be the case for fast rotating super giants (e.g. \citealt{Gvaramadze2018}, \citealt{Kourniotis2018}), the lack of signs of stellar activity in the Balmer lines and the not high enough projected rotational velocity do not favour this possibility either.

\subsection{Exocometary Break-up}

Another reasonable model for the week-long absorptions might be similar to what was proposed for the star KIC8462852 (\citealt{Boyajian2016}, \citealt{Wyatt2018}): a family of exocomets observed after breakup. We consider the possibility of an optically thick ``stream'' of ionized gas that fills out some fraction of its orbit. 

To have a better estimate of the configuration that could lead to an event lasting over several days, we tried to model the extra absorption line with a ``toy model'' of an eccentric gaseous disc. This model does not properly calculate the optical depth for different velocities but rather computes a histogram of the number of particles that have a certain velocity and are passing in the line of sight of the star.

A given model is fully described by the following parameters, the inner semi-major axis $r_0$, the radial width of the disc $\Delta r$, the eccentricity $e$, inclination $i$, position angle $\phi$, argument of periapsis $\omega$, opening angle $\psi$, stellar mass M$_\star$ stellar radius R$_\star$, and stellar radial velocity R$_\mathrm{v}$ (the latter three values being fixed to $2.15$\,M$_\odot$, $2.0$\,R$_\odot$, and $23.5$\,km.s$^{-1}$, M$_\star$ and R$_\star$ taken from \citealt{PecautMamajek2013}\footnote{\url{http://www.pas.rochester.edu/~emamajek/EEM\_dwarf\_UBVIJHK\_colors\_Teff.txt}} and R$_\mathrm{v}$ from \citealt{Iglesias2018}). We then draw $2\,000\,000$ particles, and each particle has the following orbital parameters; a semi-major axis $r$ drawn uniformly between $r_0$ and $r_0 + \Delta r$, an inclination (and position angle) drawn from a normal distribution centered at $i$ (and $\phi$, respectively) with a standard deviation of $\psi$, with all the particles sharing the same argument of periapsis $\omega$. Then, we take the mean anomaly uniformly between $[0, 2\pi)$, solve Kepler's equation for the eccentric anomaly and compute the true anomaly $\nu$ from it. We can then compute the positions and velocities $(x, y, z, v_\mathrm{x}, v_\mathrm{y}, v_\mathrm{z})$ of each particle ($z$ being the direction towards the observer). The code checks which one is passing in the line of sight of the star ($\sqrt{x^2+y^2} \leq$ R$_\star$ and $z>0$) and saves its radial velocity $v_\mathrm{z}$. We compute the histogram $F_\mathrm{mod}$ of all $v_\mathrm{z}$ (with the same binning as the observations) that satisfied the aforementioned criteria and try to fit it to the observations $F_\mathrm{obs}$. We first compute the scaling factor that minimizes the $\chi^2$, as
\begin{equation}\label{eqn:minimize}
    f_{\mathrm{scale}} = \cfrac{\sum \left( \cfrac{F_\mathrm{mod} \times (F_\mathrm{obs}-1)}{\sigma^2} \right)}{\sum \left(\cfrac{F_\mathrm{mod}}{\sigma}\right)^2}.
\end{equation}
We subtract $-1$ from $F_\mathrm{obs}$ because the observations are normalized to $1$ while the histogram starts at $0$. The final model is then obtained as $F_\mathrm{mod} = 1 + F_\mathrm{mod}\times f_\mathrm{scale}$. As already mentioned, this approach really is a ``toy model'' as we do not compute the optical depth in the photospheric line as a function of the radial velocity. Therefore we cannot really constrain the vertical extent of the disc for instance, but this allows us to have a first order approximation of where the gas should be to reproduce the velocity dispersion that we observe around HD\,37306.

\begin{figure}
	\includegraphics[width=\columnwidth]{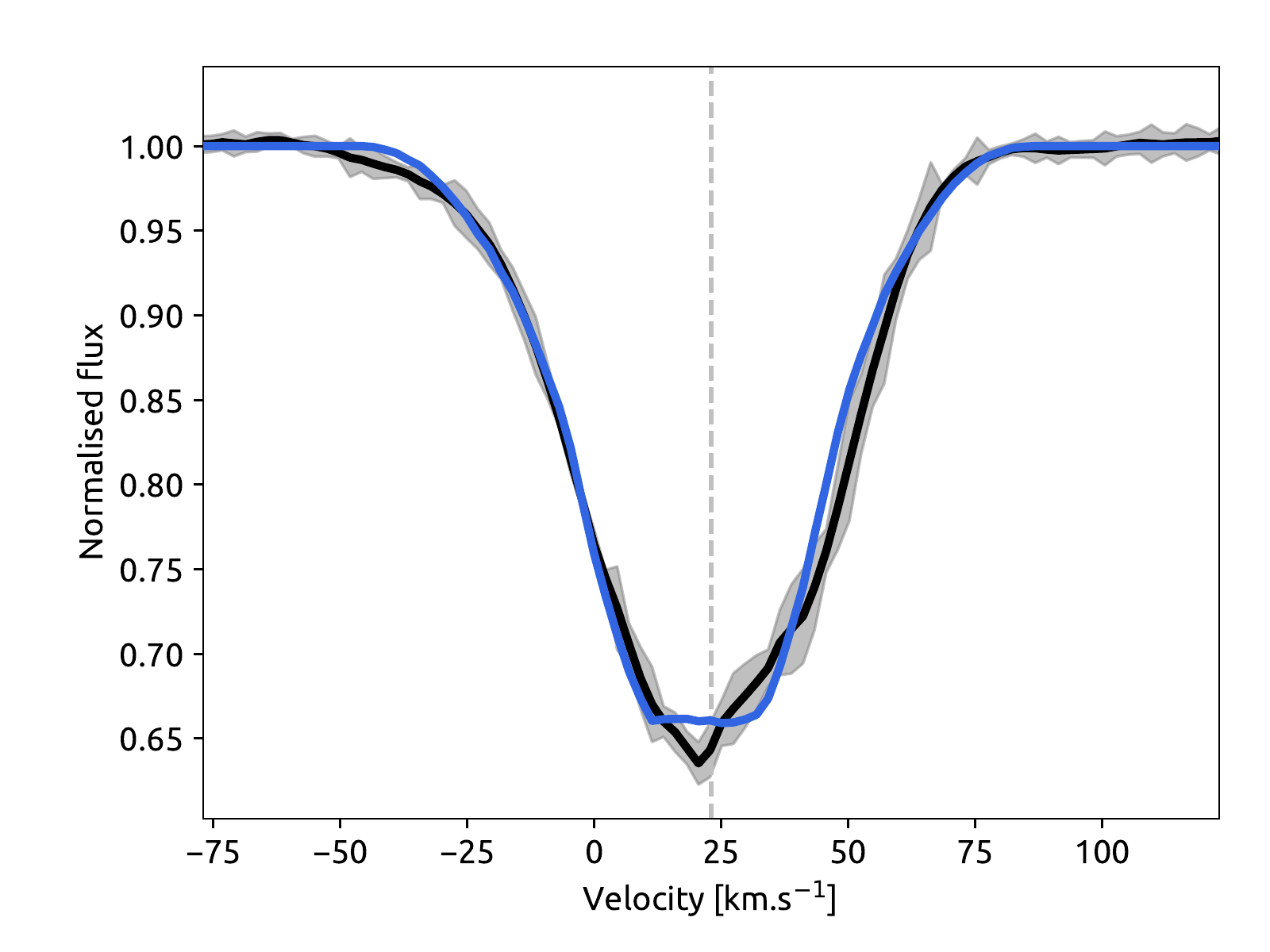}
    \caption{Model of the exo-cometary break-up (blue) to the Ca\,{\sc ii} K line absorption during the event (black). }
    \label{fig:model}
\end{figure}

\begin{figure}
	\includegraphics[width=\columnwidth]{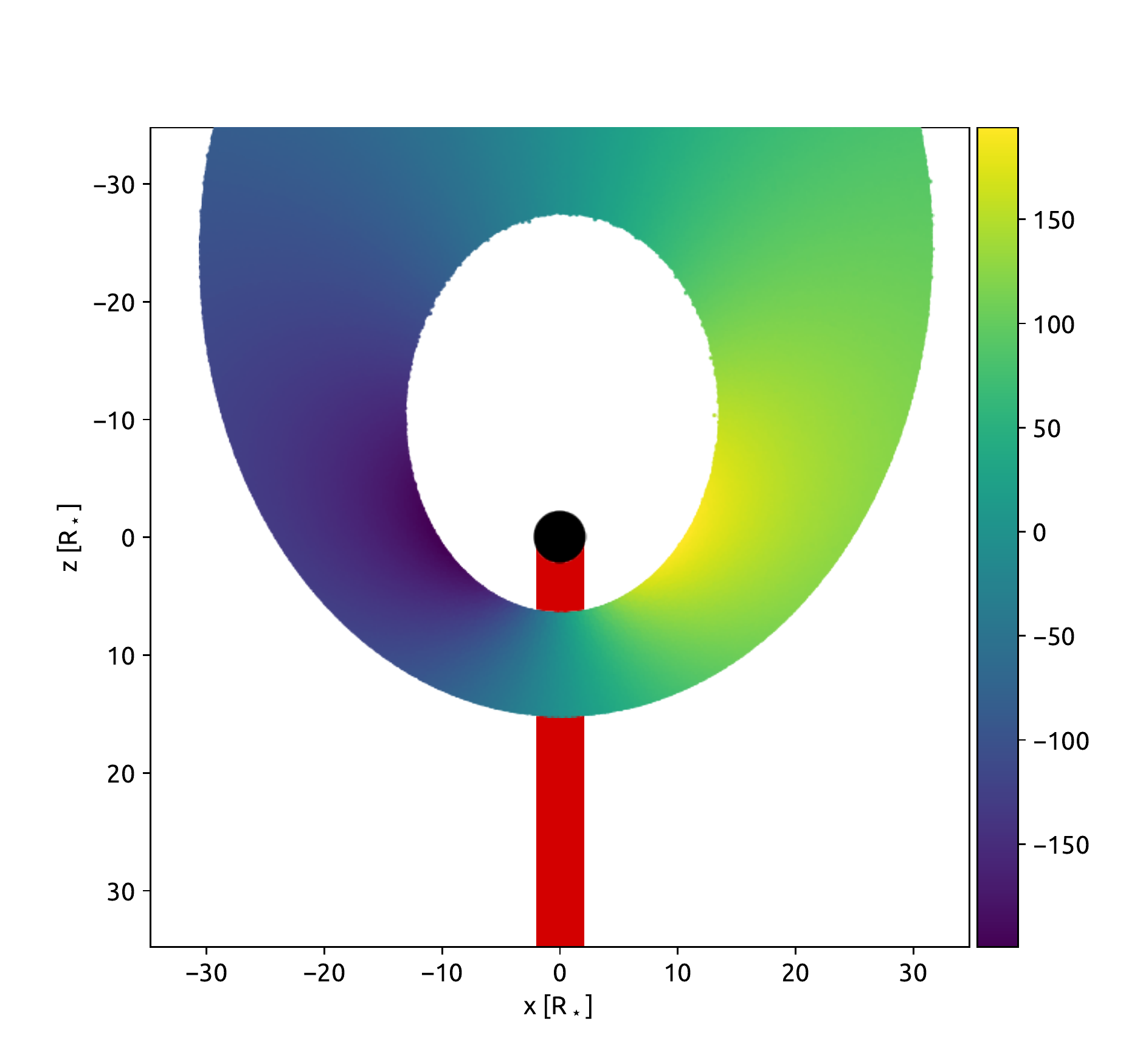}
    \caption{Top view of the best fit model. Observer located at the bottom. The color scale shows the radial velocity and the red shaded area shows where we estimate the velocity dispersion to fit the observations (the radial velocity of the star has been subtracted in this plot).}
    \label{fig:topview}
\end{figure}

\begin{figure*}
	\includegraphics[width=\textwidth]{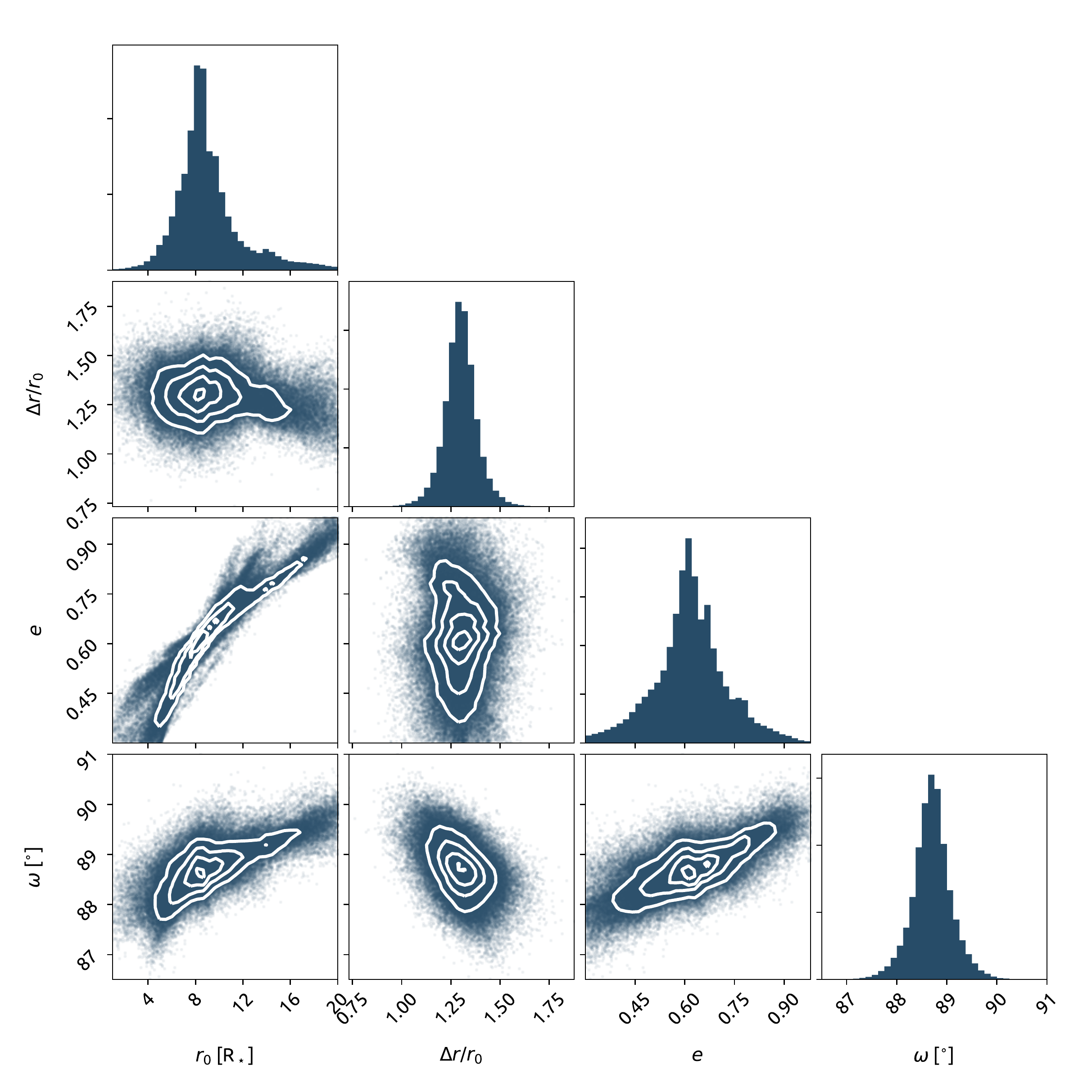}
    \caption{Probability distribution for the modeling results of the gaseous disc. }
    \label{fig:mcmc}
\end{figure*}

The problem is highly degenerate, and after several tests we settled on the following free parameters: the inner radius $r_0$, the ratio between the width of the disc and the inner radius $\Delta r/r_0$, the eccentricity $e$, and the argument of periapsis $\omega$. The value of the position angle does not matter for the modeling, and we fix $i = 90^{\circ}$ and $\psi = 0.05$. To find the best fit solution, we use an affine invariant ensemble sampler Monte-Carlo Markov Chain, implemented in the \texttt{emcee} package (\citealp{emcee2013}), using $100$ walkers, a burn-in phase of $1\,000$ steps and a final length of $2\,500$ steps. At the end of the run, we find an acceptance fraction of $0.26$ and the maximum auto-correlation length among the four free parameters is $43$ steps. Figure\,\ref{fig:model} shows the best fit model to the observations, Figure\,\ref{fig:topview} shows the top view of the same model, and the probability distributions are shown in Figure\,\ref{fig:mcmc} (using the \texttt{corner} package, \citealp{corner}). From those probability distributions we derive the $16^{\mathrm{th}}$ and $84^{\mathrm{th}}$ percentiles to estimate the uncertainties, which are reported in Table\,\ref{tab:grid}. From Figure\,\ref{fig:mcmc} it is clear that the inner edge of the disc ($r_0$) and the eccentricity of the disc ($e$) are coupled, as expected, given that we are only modeling the radial velocity dispersion. Furthermore, one should note that because the input parameter is $\Delta r/r_0$ (and not simply $\Delta r$), there is also a degeneracy between the latter value and $r_0$ (and hence with $e$ as well). We find that the gaseous disc should have inner and outer edges with semi-major axis between $8.5$ and $19.6$\,R$_\star$, and hence should be quite extended in the radial direction. Those values yield orbital periods between $5.5_{-1.6}^{+2.7}$ and $19.4_{-5.1}^{+5.6}$ days, suggesting that overall, the gaseous disc would have to survive between half and one full orbit. Considering that the chances of detection increase for longer periods and that the gas does not present significant variations over the 8 days it was observed, it is more likely that the disc should have survived close to one full orbit. The eccentricity should be around $0.6$, and we find that the pericenter of the disc should lie in between the star and the observer. This is because this configuration produces the largest radial velocity dispersion, and we do not find exactly $90^{\circ}$ because of the slight asymmetry of the profile. If we consider the inclination of the orbiting material is close to $90^{\circ}$ and that the lower limit on the stellar fraction covered by the gas is $\sim$37\%, then a lower limit on the opening angle of the disc should lie between 0.015 and 0.035, at the outer and inner edges of the disc, respectively. We should note that the model ignores the possible effect of radiation pressure, which might accelerate the dissipation of the gas. 

\begin{table}
\caption{Free parameters, priors, and best fit results for the modeling of the observations assuming a gaseous disc around HD\,37306.}
\label{tab:grid}
\centering
\begin{tabular}{lcc}
\hline\hline
Parameters & Prior & Best-fit \\
\hline
$r_0$ [R$_\star$]     & [$1$, $20$]      & $8.5_{-1.7}^{+2.6}$ \\
$\Delta r / r_0$      & [$0.1$, $5$]     & $1.3_{-0.1}^{+0.1}$ \\
$e$                   & [$0.3$, $0.98$]  & $0.6_{-0.1}^{+0.1}$ \\
$\omega$ [$^{\circ}$] & [$80$, $100$]    & $88.7_{-0.3}^{+0.3}$ \\
\hline
\end{tabular}
\end{table}

Because the vast majority of the $2\,000\,000$ particles do not pass in front of the star, the final modeled spectrum can be relatively noisy. Instead of increasing significantly the number of particles we smooth the modeled spectrum using a moving box; each velocity bin is the mean value of the $3$ neighbours before and $3$ neighbours after (including the bin that is evaluated). Because the observed spectrum is also noisy, especially close to the minimum, we also smooth it in a similar fashion.

As can be appreciated in Fig.\,\ref{fig:model}, the model reproduces the absorption profile quite closely and the scenario is compatible with the duration of the event. In addition, the lack of detection in the simultaneous photometric observations taken with MASCARA and ASAS-SN could be explained by not reaching the sensitivity required to detect an exocomet transiting an A-type star. For instance, the first photometric detection of exocomets crossing in front of $\beta$ Pictoris (also an A-type star) were performed using TESS observations and the dips were reported to have depths between 0.5 and 2 millimagnitudes \citep{Zieba2019}. Between MASCARA and ASAS-SN, MASCARA reached the best sensitivity with $\sigma$=17.7 millimagnitudes, clearly not sensitive enough to have a detection similar to those in $\beta$ Pictoris. Unfortunately, the available TESS observations of HD\,37306 were not simultaneous with the spectroscopic event.

\subsection{Colliding Trojans}

 We studied the possibility of such a long lasting absorption being due to colliding trojans producing gas in the Lagrangian points of a possible hidden planet. But, in order to be so close to the star and transit for so long, the orbit of the planet needs to be highly eccentric. We performed hydrodynamical simulations of a gaseous disc with an embedded Jupiter mass planet using our modified version \citep{Montesinos+2015} of the Fargo code \citep{Baruteau-Masset-2008}. The surface disc density is defined by  $\Sigma(r)  = \Sigma_0 \frac{[\rm 1 au] }{r}$, where $\Sigma_0 = 0.1 \rm{~ g.cm^{-2}}$, and  $r$ the distance to the star in au. We assume a viscous fluid with a turbulent visocosity \citep{Shakura+Sunyaev1973} given by $\alpha$ = $10^{-3}$. Under these circumstances, most of the gas is accreted during the simulation ($10^{5}$ yrs), remaining an optically thin layer of gas with a gap (carved by the planet). The resulting low density disc could be compatible with the lack of detection of a stable gas component. The quantification of the lack of detection is, however, out of the scope of this paper, since stability arguments discard this scenario (see below). On top of that simulation, we followed the evolution of (100,000) dust particles with different sizes ranging from microns to cms, where the dust is affected by the gas dynamics due to drag and drift forces.  In addition to these forces, the dust particles feel the gravitational potential from the star. The self-gravity of the disc was neglected (details of our dust code can be found in \citealp{Cuello2019}).
 
We found that the presence of a Jupiter mass planet in a circular orbit carves a gap in the gaseous disc with a large dust particle concentrations ($\sim 1$ Earth-mass) located at the Lagrangian points (L4 and L5) of the giant planet (in agreement with \citealp{Lyra+2009}). In this case, the dusty trojans rotate at Keplerian velocity with zero eccentricity. For Jupiter mass planets with eccentricities higher than 0.1, we found that the dust is not able to accumulate in the Lagrangian points. It is worth mentioning that in order to explain the observations, trojan eccentricities should be at least 0.6.

\subsection{Planetary Transit}

Planetary transits are very difficult to detect in A-type stars using photometry given the brightness of the star compared to the shadow produced by a planet. However, the brighter the star, the higher the signal to noise of the transiting planet atmospheric spectrum. Here we propose a couple of scenarios involving a transiting planet. 



\begin{figure}
	\includegraphics[width=\columnwidth]{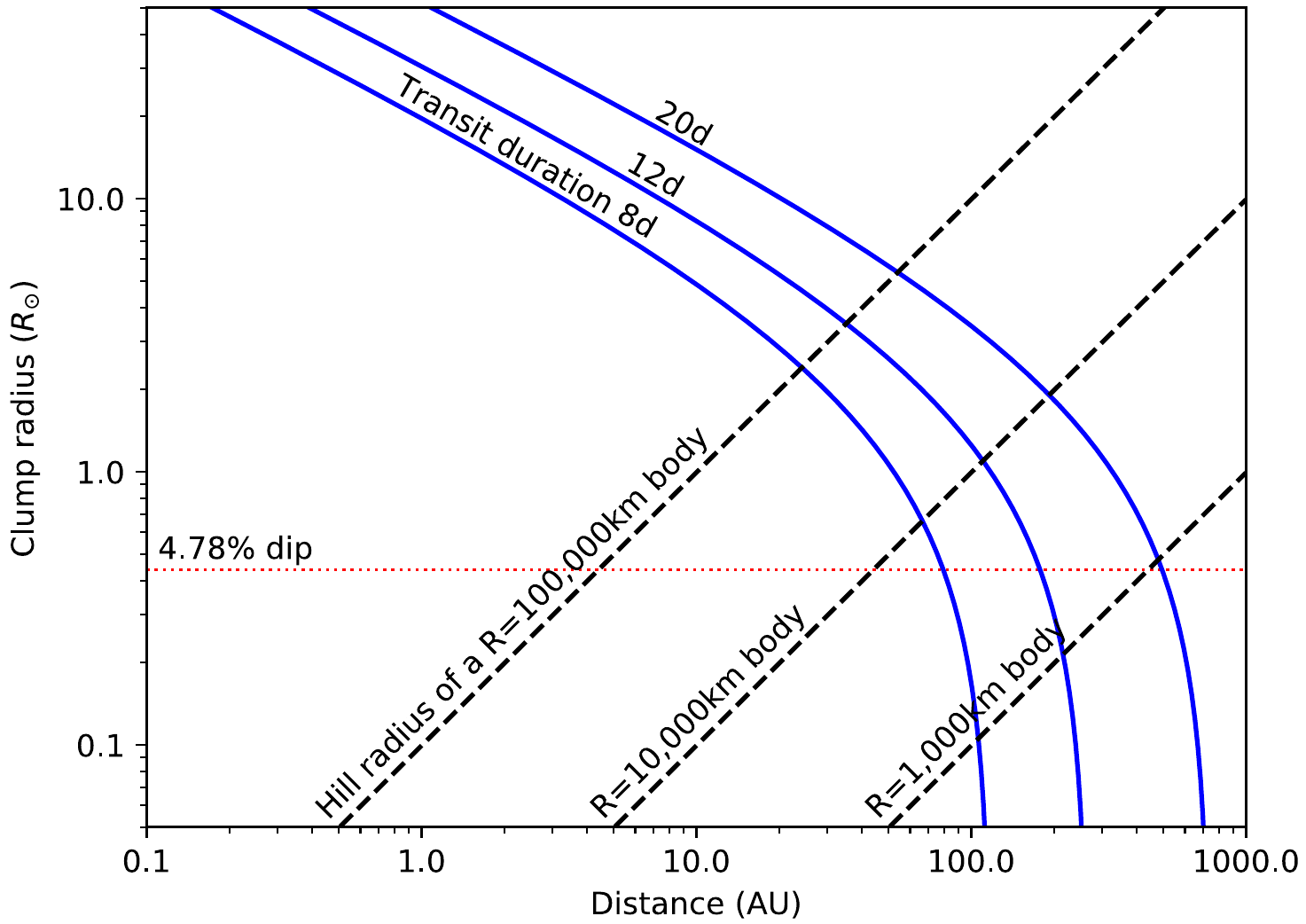}
    \caption{Distance to the star versus size of an optically thick, spherical dusty clump on circular orbits around the star. The solid blue lines show the transit duration, diagonal black dashed lines show Hill radii of planets of different sizes, and the red dotted line shows the minimum size the clump should have in order to have a $3\sigma$ detection in the MASCARA photometry, which corresponds to a 4.78\% dip in the light curve.}
    \label{fig:scenario}
\end{figure}

A possible scenario could invoke a planet passing at close enough distance to the star allowing sublimation of its atmosphere (or even surface) to occur. To better understand the possible location and size of what could be causing the observed absorptions, we explore the scenario-independent constraints, similar to the analysis done in \cite{Boyajian2016}. Fig. \ref{fig:scenario} (analog to Fig. 10 in \citealt{Boyajian2016}) shows constraints for the duration of the transit, assuming that the clump causing the dip is opaque and moves in circular orbits around the star. As mentioned in previous sections, for the stellar parameters we assume $R_{\star}=2.0R_{\odot}$ and $M_{\star}=2.15M_{\odot}$ \citep{PecautMamajek2013}. Following \cite{Boyajian2016}, if the clump is much less massive than the star of mass $M_{\star}$ and orbits at a distance $d$ from the star, then the relation between the clump radius $r$ and the duration of the transit $t$ is given by,

\begin{equation}\label{eqn:transitDurat}
    r\approx 1.85 t\sqrt{\cfrac{M_{\star}}d} - R_{\star}.
\end{equation}

In this scenario, we are assuming the clumps to be spherical and, basically, formed by a planet (or planetesimal) and the accumulation of sublimating material (gas and dust) inside its Hill sphere of radius $R_{\mathrm{Hill}}=d(M_{\mathrm{pl}}/[3M_{\star}])^{1/3}$.

Having no significant dip detected in the light curves ($>3\sigma$), the clump size must be under this detection limit (dotted red line in Fig. \ref{fig:scenario}), which, considering $R_{\star}=2.0R_{\odot}$, translates into $r\approx0.44R_{\odot}$. In order for the duration of the transit to be of at least eight days and be under the photometric detection limit, the clump should orbit at a distance $\geq\sim 100$\,au. At this distance, the temperature would be too low for the planet to sublimate producing large dense clouds around it, and for the gas to ionize. In addition, if the probability of having a planetary transit is $\sim  R_{\star}/d$, then at $\sim100$\,au this probability would be of $9.3 \times 10^{-5}$, making the probability of the detection very low.

We could also consider the case of a forming planet having a Hill sphere full of gas. At the very early phase of planet formation, the planet has an intrinsic luminosity due to accretion (peaking in L, L' bands), and what we would see would be actually the circumplanetary disc (not the planet itself). In this case, a forming (accreting) planet should be very bright (one should expect $10^{-3}$ -- $10^{-6}$ $L_{\odot}$ during the first million years of formation; \citealt{Mordasini2012}). This scenario only makes sense in a gas-rich environment, therefore this is not likely to be the right explanation taking into account the debris disc nature and age range of the system. 
In addition, as discussed in the first case, to explain an eight days transit the forming planet would have to be far away from the star but, at the same time, the gas would have to be ionized, making the second possible explanation in this section very unlikely as well.


\subsection{Possible interstellar origin}

In the diffuse interstellar medium (ISM), Ca\,{\sc ii} and Ti\,{\sc ii} are pretty well correlated \citep{Hunter2006} and the Na\,{\sc i}/Ca\,{\sc ii} ratio ranges from $0.1$ to $1000$ (\citealt{Siluk+Silk1974}, \citealt{Vallerga1993}) depending on the dust and ionization state of the Ca\,{\sc ii}, which is not in disagreement with our detection. However, the median value of the $b$ value 
for Ca\,{\sc ii} components is $1.3$\,km.s$^{-1}$ in the diffuse ISM, corresponding to a kinetic temperature of around $4100$\,K \citep{Welty1996}. Our $b$ values for the Ca\,{\sc ii} components are of $33.2$\,km.s$^{-1}$ on average, which would correspond either to very high turbulent velocities ($>20$\,km.s$^{-1}$) or extremely high kinetic temperatures (> $2 \times 10^{6}$\,K). More importantly, interstellar features do not show such strong variations within short periods (e.g. \citealt{Smith2013} and \citealt{McEvoy2015}), thus the appearance and disappearance of a large component within three years would be very unlikely. Besides, the star is at $\sim 70$\,pc, thus it is located inside the local bubble where column densities do not reach such high values. For instance, at 100pc $\log N$(Ca\,{\sc ii}) $\sim 11.5$\,cm$^{-2}$ \citep{Hunter2006}, and the observed absorptions have on average a $\log N$(Ca\,{\sc ii}) of $\sim 12.9$\,cm$^{-2}$.

\section{Conclusions}


We detected unusually broad spectral absorption features from ionized gas species in the debris disc system HD\,37306. The stellar spectra present stable non-photospheric absorption lines superimposed onto several Ca\,{\sc ii}, Fe\,{\sc ii} and Ti\,{\sc ii} lines over a time range of at least eight days.

We analysed simultaneous spectroscopic and photometric observations of HD\,37306 before, during and after the event and found no significant change in the photometric data that might be correlated with the spectroscopic event. We also analysed high angular resolution SPHERE images of the target but the disc was not detected.

We evaluated several scenarios aiming to determine which one would provide the most likely explanation for the particular features detected in HD\,37306. We were able to reasonably discard some possibilities such as an instrumental artifact, colliding trojans or an interstellar origin. Other options such as a circumstellar shell or a planetary transit cannot be completely ruled out, but remain as very unlikely possibilities. 

We conclude that the most likely scenario would be the transit of a family of exocomets observed after breakup releasing a large amount of gas close to the star, at a few stellar radii. Our model reproduces satisfactorily the broad profile of the absorption feature and is in agreement with the duration of the event. A rather strong limitation of this scenario is that radiation pressure should dissipate Ca\,{\sc ii} on short timescales and this effect is not considered in our model.

Due to the stochastic nature of the event we reported in this study, it will be challenging to further characterize what happened during 2017. Future ALMA observations may help detect other gas tracers in emission, such as CO, that could have been released during this event, and may help us figure out the processes involved in this unusual week-long event. We will perform further high resolution spectroscopic observations of this system aiming to possibly detect any kind of additional events and characterize them to have a better understanding of the system. Given the rarity of this event, we aim to obtain an estimate of the occurrence rate of such phenomenon in debris disc stars.

\section*{Acknowledgements}


AB, DI, JO, MM and NG acknowledge support from ICM (Iniciativa Cient\'ifica Milenio) via N\'ucleo Milenio de Formaci\'on Planetaria. A.B. acknowledges support from FONDECYT regular grant 1190748.
JO acknowledges financial support from the Universidad de Valpara\'iso, and from Fondecyt (grant 1180395). M.M. acknowledges financial support from the Chinese Academy of Sciences (CAS) through a CAS-CONICYT Postdoctoral Fellowship administered by the CAS South America Center for Astronomy (CASSACA) in Santiago, Chile. GMK is supported by the Royal Society as a Royal Society University Research Fellow. NG acknowledges grant support from project CONICYT-PFCHA/Doctorado Nacional/2017 folio 21170650. Based on observations collected at the European Southern Observatory under ESO programmes: 076.C-0279(A), 078.C-0209(A), 080.C-0712(A), 094.A-9012(A), 099.A-9029(A), and 095.C-0212(B). DI thanks Isabel Rebollido and Dr. Benjam\'in Montesinos for their insightful comments on this paper.
This paper includes data collected by the TESS mission, which are publicly available from the Mikulski Archive for Space Telescopes (MAST). Funding for the TESS mission is provided by the NASA Explorer Program. This research has made use of the SIMBAD database, operated at CDS, Strasbourg, France, and Astropy, a community-developed core Python package for Astronomy \citep{Astropy2013}. This work has made use of data from the ESA space mission Gaia (http://www.cosmos.esa.int/gaia), processed by the Gaia Data Processing and Analysis Consortium (DPAC, http://www.cosmos.esa.int/web/gaia/dpac/consortium). Funding for the DPAC has been provided by national institutions, in particular the institutions participating in the Gaia Multilateral Agreement. This publication makes use of VOSA, developed under the Spanish
Virtual Observatory project supported from the Spanish MICINN through grant AyA2011-24052.















\bsp	
\label{lastpage}
\end{document}